\documentclass[12pt]{article}
\usepackage{amssymb}
\usepackage{amsfonts}
\newcommand{\eqd}{\stackrel{\rm def}{=}}%
\newcommand{\normII}[1]{\mathop{\rm vrai\,sup}\limits_{(q,q')
\in {\cal V}^{(2)}} \left|{#1}\right|}%

\begin{document}

\author{M. I.~Kalinin\thanks{The Russian Research Institute for
Metrological Service},\\
46 Ozernaya str., Moscow, Russia, 119361. \quad kalinin@vniims.ru}
\title{On one-to-one correspondence of Gibbs distribution and
reduced two-particle distribution function}
\date{}
\maketitle

\thispagestyle{empty}

\begin{abstract}
In this article it is shown that in an equilibrium classical
canonical ensemble of molecules with two-body interaction and
external field full Gibbs distribution can be uniquely expressed
in terms of a reduced two-particle distribution function. This
means that while a number of particles $N$ and a volume $V$ are
fixed the reduced two-particle distribution function contains as
much information about the equilibrium system as the whole
canonical distribution. The latter is represented as an absolutely
convergent power series relative to the reduced two-particle
distribution function. As an example a linear term of this
expansion is calculated. It is also shown that Gibbs distribution
function can de expressed in terms of reduced distribution
function of the first order and pair correlation function.That is
the later two functions contain the whole information about system
under consideration.
\end{abstract}

\section{Introduction}

In classical statistical mechanics an equilibrium system of $N$
molecules in a volume $V$ is described by canonical distribution
function $F_N(q, p)$, where $(q, p)$ is a set of phase variables:
coordinates $q_i$ and momenta $p_i$ of molecules. If interaction
of molecules is additive, reduced distribution functions are
introduced \cite{NN,Balescu}. They are used for evaluation of
thermodynamic characteristics of this molecular system. It's
usually accepted that reduced distribution functions contain
information about a molecular system less than the initial
canonical distribution function. It's also supposed that the lower
an order of a reduced distribution function is, the less
information it contains. But there does not exist a proof of this
statement in scientific literature.

On the other hand, it is known that for an equilibrium canonical
ensemble of non-interacting particles  a canonical distribution
function $F_N(q,p)$ is decomposed into a product of reduced
one-particle distribution functions $F_1(q,p)$ \cite{NN,Balescu}.
This means that all information about such system is contained in
the reduced one-particle distribution function.

In the article \cite{Kalinin} it was proved that for a system with
pair interaction and without external field a reduced two-particle
distribution function defined in terms of nonormalized Gibbs
distribution (without configurational integral) contains the whole
information about such system. In \cite{Kalinin2} that result was
obtained for normalized Gibbs distribution and respective reduced
distribution function. This results were obtained for a system
which is not subjected to influence of external field.

In this paper it's proved that for such system placed into
external field one-to-one correspondence between full canonical
distribution function and a reduced two-particle distribution
function exists. Moreover it's shown that one-to-one
correspondence between a full canonical Gibbs distribution and a
set of reduced distribution function of the first order together
with pair correlation function exists too. This means that the
reduced two-particle distribution function as well as mentioned
set contain the total information about system under
consideration.

We consider an equilibrium system of $N$ particles contained in
the volume $V$ under the temperature $T$. Potential energy of
system is supposed to have the form
\begin{equation}
\label{U_N} %
U_N(q_1,\ldots,q_N)=\sum_{i=1}^Nu_1(q_i)+\sum_{1\le j<k\le
N}u_2(q_j,q_k),
\end{equation}                                            
where $u_1(q)$ is an external field and $u_2(q,q')$ is a two-body
interaction energy of particles. Probability distribution function
of equilibrium system is the canonical Gibbs distribution which is
decomposed into a product of a momentum distribution function and
a configurational one \cite{NN,Balescu}. The former is expressed
as a product of one-particle Maxwell distributions, the latter has
the form
\begin{equation}
\label{D_N} %
D_N(q_1,\ldots,q_N)=Q_N^{-1}\exp\{-\beta U_N(q_1,\ldots,q_N)\},
\end{equation}                                            
where $\beta=1/kT$, $k$ is the Boltzmann constant and $Q_N$ is the
configuration integral
\begin{equation}
\label{Q_N} %
Q_N=\int\!\!\exp\{-\beta U_N(q_1,\ldots,q_N)\}dq_1 \cdots dq_N.
\end{equation}                                            
Here and below integrating with respect to every configurational
variable is carried out over the volume $V$. For a system having
interaction of form (\ref{U_N}) reduced distribution functions are
introduced by expressions \cite{Balescu}
\begin{equation}
\label{F_l} %
F_l(q_1,\ldots,q_l)=\frac{N!}{(N-l)!}\int D_N(q_1,\ldots,q_N)
dq_{l+1}\cdots dq_N, \quad l=1,2,\ldots .
\end{equation}                                            
These functions are used instead of full canonical distribution
(\ref{D_N}) to calculate various characteristics of the molecular
system. Let us investigate properties of the reduced two-particle
distribution function.

Potential energy (\ref{U_N}) can be written as
\begin{equation}
\label{U_N1} %
U_N(q_1,\ldots,q_N)=\sum_{1\le j<k\le N}\phi(q_j,q_k),
\end{equation}                                            
where
\begin{equation}
\label{phi_s} %
\phi(q,q')=u_2(q,q')+\frac{u_1(q)+u_1(q')}{N-1}.
\end{equation}                                            

Introduce a function $h(q,q')$ by the relation
\begin{equation}
\label{h} %
\exp\{-\beta \phi(q,q')\}= \sigma\{1+h(q,q')\},
\end{equation}                                            
where
\begin{equation}
\label{sigma} %
\sigma= \frac{1}{V^2}\int\exp\{-\beta \phi(q,q')\}\,dqdq'.
\end{equation}                                            
The canonical Gibbs distribution (\ref{D_N}) takes the form
\begin{equation}
\label{D_N1} %
D_N(q_1,\ldots,q_N)=Q_N^{-1}\prod_{1\le j<k\le N}[1+h(q_j,q_k)]
\end{equation}                                            
with
\begin{equation}
\label{Q_N1} %
Q_N=\int\!\!\prod_{1\le j<k\le N} [1+h(q_j,q_k)]\,dq_1 \cdots
dq_N.
\end{equation}                                            

From (\ref{D_N1}) and (\ref{Q_N1}) it follows that statistical
properties of system under consideration are completely determined
by the specifying single function of two configurational variables
$h(q,q')$ and two external parameters $N$ and $V$. Gibbs
distribution (\ref{D_N1}) has the same form as for system without
external field. Therefore we may suppose that results of article
\cite{Kalinin2} can be proved for system considered in this
article too.

In section \ref{problem} we state a mathematical problem for our
molecular system  and formulate conditions for existence and
uniqueness of its solution. In section \ref{proof} feasibility of
these conditions for considered physical system are proved. In
section \ref{sec-h(f)} an expression for function $h(q,q')$ in
terms of $F_2(q,q')$ is calculated. In section \ref{sec-phi(f)} an
expression for the canonical distribution in terms of reduced
reduced two-particle distribution function is produced. In section
\ref{set} it is shown that a set of first reduced distribution
function and pair correlation function contains the whole
information as well as two-particle distribution function and
canonical Gibbs distribution.

\section{Mathematical formulation of problem\label{problem}}

Let us introduce function $f(q,q')$ by the relation
\begin{equation}
\label{f} %
F_2(q,q')= \frac{N!}{(N-2)!\,V^2}[1+f(q,q')].
\end{equation}                                            
 Both the function $f(q,q')$ and the function $h(q,q')$
 satisfy the conditions
\begin{equation}
\label{cond_f-h} %
\int f(q,q')\,dqdq'=0, \qquad \int h(q,q')\,dqdq'=0.
\end{equation}                                       

From expressions (\ref{F_l}), (\ref{D_N1}), and (\ref{f}) it
follows that
\begin{equation}
\label{h-f} %
1+f(q_1,q_2)=\frac{V^2}{Q_N}\int\prod_{1\le j<k\le N}
[1+h(q_j,q_k)]dq_3\cdots dq_N.
\end{equation}                                            
This relation defines the transformation $\{h\rightarrow f\}$ and
can be considered as a nonlinear equation relative to $h(q,q')$.
If there exists a solution $h(q,q';[f])$ of this equation then a
function $D_N(q_1,\ldots,q_N)$ becomes an operator function of
$f$. It means that both the canonical Gibbs distribution $D_N$ and
all reduced distribution functions $F_l$ are expressed in terms of
a single reduced distribution function $F_2$. Thus we have to
prove that equation (\ref{h-f}) has a unique solution and
therefore the transformation $\{h\rightarrow f\}$  has inverse one
$\{f\rightarrow h\}$.

Multiplying equation (\ref{h-f}) by $Q_NV^{-N}$ and using
(\ref{Q_N1}) we rewrite it in the form
\begin{eqnarray}
 [1+f(q_1,q_2)]\frac{1}{V^N}\int\!\!\prod_{1\le j<k\le N}
 [1+h(p_j,p_k)]\,dp_1 \cdots dp_N    \nonumber\\
-\frac{1}{V^{N-2}}\int\prod_{1\le j<k\le N}
[1+h(q_j,q_k)]dq_3\cdots dq_N=0.
\label{h-f_1} %
\end{eqnarray}                                            
The left-hand side of (\ref{h-f_1}) is a polynomial operator of
degree ${\cal N}=N(N-1)/2$ relative to $h$ and of degree one
relative to $f$. Denote this operator by ${\cal F}(h,f)$. Equation
(\ref{h-f_1}) can be written in a symbolic form
\begin{equation}
\label{F(h-f_2)} %
{\cal F}(h,f)=0.
\end{equation}                                            
To solve this equation it is necessary to specify an additional
condition
\begin{equation}
\label{F(h0,f0)} %
{\cal F}(h^{(0)},f^{(0)})=0,
\end{equation}                                            
where $f^{(0)}(q,q')$ and $h^{(0)}(q,q')$ are assigned functions.

We can easily determine these functions for our physical system.
If the external field and interaction between particles are
absent, i.e., potentials $u_1$ and $u_2$ are constant, then the
function $h(q,q')$ vanishes. Under this condition
$Q_N=Q_N^{(0)}=V^N$, $D_N(q_1,\ldots,q_N)=D_N^{(0)}=V^{-N}$, and
$f(q,q')=0$. Therefore we can take $h^{(0)}=0$ and $f^{(0)}=0$ in
(\ref{F(h0,f0)}).

Equation (\ref{F(h-f_2)}) and additional condition
(\ref{F(h0,f0)}) form a problem on implicit function. In
functional analysis there is a number of theorems on implicit
function for operators of various smoothness classes. We use the
theorem for analytic operator in Banach space in the form given in
the book \cite{Vainberg}.

\medskip %
{\bf Theorem.} (On implicit function). {\it Let ${\cal F}(h,f)$ be
an analytic operator in $D_r(h^{(0)},E_1)\times
D_{\rho}(f^{(0)},E)$ with values in $E_2$. Let an operator $B\eqd
-\partial {\cal F}(h^{(0)},f^{(0)})/\partial h$ have a bounded
inverse one. Then there are positive numbers $\rho_1$ and $r_1$
such that a unique solution $h=\chi(f)$ of the equation ${\cal
F}(h, f)=0$ with the additional condition ${\cal F}(h^{(0)},
f^{(0)})=0$ exists in a solid sphere $D_{r_1}(h^{(0)},E_1)$. This
solution is defined in a solid sphere $D_{\rho_1}(f^{(0)},E)$, is
analytic there, and satisfies the condition
$h^{(0)}=\chi(f^{(0)})$}.

\medskip %
\noindent %
Here $D_r(x_0,{\cal E})$ denotes a solid sphere of radius $r$ in a
neighborhood of the element $x_0$ in a normalized space ${\cal
E}$, the symbol $\times$ denotes the Cartesian product of sets,
$\partial {\cal F}/\partial h$ is Fr\'{e}chet derivative
\cite{Vainberg,Dunford} of the operator ${\cal F}$, $h^{(0)}$ and
$f^{(0)}$ are assigned elements of the respective spaces $E_1$ and
$E$. If the functions $h(q,q')$ and $f(q,q')$ satisfy this theorem
conditions, the former is a single valued operator function of the
later.

To prove an existence and uniqueness of the solution of problem
(\ref{F(h-f_2)}), (\ref{F(h0,f0)}) we have to show that conditions
of the above theorem are satisfied.

\section{Proof of feasibility of the theorem conditions \label{proof}}

First we define spaces $E_1$, $E$, and $E_2$ mentioned in the
theorem for functions describing the physical system under
consideration.

\subsection{Functional spaces of problem}

Potentials $u_1(q)$ and $u_2(q,q')$ are real symmetric functions.
Suppose they are bounded below for almost all $\{q,q'\}\in V$. All
physically significant potentials possess this property. Under
this condition integral (\ref{sigma}) exists and $h(q,q')$ is a
real symmetric function bounded for almost all $\{q,q'\}\in V$.

A set of functions bounded nearly everywhere forms a complete
linear normalized space (Banach space) with respect to the norm
\cite{Dunford,Kantorovich}
\begin{equation}
\label{norm_h} %
\|h\|=\normII{h(q,q')},
\end{equation}                                          
where "$\mathop{\rm vrai\:sup}$" denotes an essential upper bound
of the function on the indicated set and ${\cal V}^{(2)}\eqd
V\mathop{\times}V$ is Cartesian product of $V$ by itself. It is
called the space of essentially bounded functions and is denoted
by $L_{\infty}({\cal V}^{(s)})$. In addition $h(q,q')$ satisfies
condition (\ref{cond_f-h}). The set of such functions is a
subspace of $L_{\infty}({\cal V}^{(2)})$. It is easy to show that
this subspace is a complete space relative to norm (\ref{norm_h}).
Therefore we can take the Banach space of symmetric essentially
bounded functions satisfying condition (\ref{cond_f-h}) as $E_1$.

Expression (\ref{h-f}) for $f(q,q')$ includes multiple integrals
of different power combinations of $h(q,q')$. Any power of
essentially bounded function are integrable with respect to
arbitrary set of variables $\{q_{j_1},\ldots,q_{j_r}\}$ over $V$
\cite{Vulikh}. Therefore all integrals in (\ref{h-f_1}) are
essentially bounded functions too. Arguing as above, we can show
that the space $E$ of functions $f(q,q')$ coincides with $E_1$.
Continuing in the same way we can show from (\ref{h-f_1}) and
(\ref{F(h-f_2)}) that $E_2$ is the same space. Thus we define the
spaces of the above theorem as $E=E_1=E_2=L_{\infty}({\cal
V}^{(2)})$ with property (\ref{cond_f-h}).

From (\ref{h}) and (\ref{f}) for $h(q,q')$ and $f(q,q')$ it
follows that $f>-1$ and $h>-1$. Therefore we can take a manifold
\{$f>-1, \; h>-1$\} as a definition domain of the operator ${\cal
F}(h,f)$. Since the left-hand side of (\ref{h-f_1}) is a
polynomial, the operator ${\cal F}(h,f)$ is analytical in this
domain. As stated above the additional condition (\ref{F(h0,f0)})
is valid for $h^{(0)}=f^{(0)}=0$. Thus any solid spheres of $E_1$,
$E$ with centers at $h^{(0)}=0$, $f^{(0)}=0$ and radii $r<1$,
$\rho<1$ respectively can be used as domains $D_r(h^{(0)},E_1)$
and $D_{\rho}(f^{(0)},E)$ indicated in the theorem.

Finally it is necessary to prove that the operator
\begin{equation}
\label{B} %
\left. B\eqd -\frac{\partial {\cal F}(h,f)}{\partial
h}\right|_{{h=0,\atop f=0\phantom{,}}\atop \vphantom{0}}
\end{equation}                                          
has a bounded inverse one.

\subsection{Properties of the operator $B$}

To find the inverse operator $B^{-1}$ it is necessary to solve the
equation $Bh=y$, where $h\in E_1$, $y\in E_2$. The expression for
$Bh$ is a linear relative to $h$ part in the left-hand side of
relation (\ref{h-f_1}) as $f=0$. Let us introduce next notations
for arbitrary function $\xi(q,q')$
\begin{equation}
\label{h-(k)} %
\overline{\xi}(q)=\frac{1}{V}\int \xi(q,q')dq', \qquad
\overline{\overline{\xi}}=\frac{1}{V^2}\int \xi(q,q')dqdq'.
\end{equation}                                            
Expanding products in (\ref{h-f_1}) and keeping linear summands we
obtain
\begin{equation}
\label{Bh} %
(Bh)(q,q'))=h(q,q')+(N-2)[\overline{h}(q)+\overline{h}(q')].
\end{equation}                                            

It's easy to estimate a norm of the operator $B$. Using definition
(\ref{norm_h}) we obtain
\begin{eqnarray}
\|Bh\|\le(2N-3)\|h\|.
\label{||Bh||} %
\end{eqnarray}                                            
From here we get an estimation
\begin{equation}
\label{||B||} %
\|B\|\le 2N-3.
\end{equation}                                            
Thus the operator $B$ is bounded.

Using (\ref{Bh}) we can write a nonuniform equation $Bh=f$ in the
next form
\begin{equation}
\label{Bh3} %
h(q,q')+(N-2)[\overline{h}(q)+\overline{h}(q')]=f(q,q').
\end{equation}                                            
Taking into account the conditions (\ref{cond_f-h}) we get a
solution of this equation
\begin{equation}
\label{h(y)} %
h(q,q)=(B^{-1}f)(q,q)=f(q,q')-\frac{N-2}{N-1}\,
[\overline{f}(q)+\overline{f}(q')].
\end{equation}                                            
From here it's easy to estimate a norm of the inverse operator
$B^{-1}$. Evaluating the norm of right-hand-side of (\ref{h(y)})
we get
\begin{equation}
\label{||B^-1||} %
\|B^{-1}\|\le \frac{\|B^{-1}f\|}{\|f\|}\le
\left|1+2\frac{N-2}{N-1}\right|=
\left|\frac{3(N-1)-2}{N-1}\right|\le 3.
\end{equation}                                        
Therefore the operator $B^{-1}$ exists and it is bounded.

So all conditions of the above theorem are valid for our physical
system. Hence there exists a unique solution $h=\chi(f)$ of
problem (\ref{F(h-f_2)}), (\ref{F(h0,f0)}) as a function of $f$.
This solution defines an inverse transformation from the function
$f$ to the function $h(q,q')=h(q,q';[f])$.

\section{Derivation of the inverse transformation $h(f)$ \label{sec-h(f)}}

To obtain the transformation $f\rightarrow h$ we have to solve
equation (\ref{h-f_1}) relative to $h(q,q')$. Here we present less
unwieldy derivation than in the work \cite{Kalinin2}. At first
define an auxiliary operator function $g(h)$ by means of a
relation
\begin{equation}
\label{g(h)-def} %
\frac{1}{V^{N-2}}\int\prod_{1\le j<k\le N}
[1+h(q_j,q_k)]dq_3\cdots dq_N=1+g(h).
\end{equation}                                       
This operator function is a polynomial of degree ${\cal N}$
relative to $h$ and depends on two configurational variables
$q_1,q_2$. It can be written in the form
\begin{equation}
\label{g(h)} %
g(h)= \sum_{l=1}^{\cal N}g_l(h),
\end{equation}                                          
where $g_l(h)$ is a uniform operator of order $l$ relative to $h$.
Let us derive the expression for $g_l(h)$ from definition
(\ref{g(h)-def}).

For the sake of abbreviation of subsequent calculations we
introduce next notations. We will denote by number
$K\in\{1,\ldots,{\cal N}\}$ every ordered collection $\{j,k\}$
from the set $\{1,\ldots,N\}$. Such one-to-one correspondence can
be always made. A collection $\{q_j,q_k\}$ is an element of
manifold ${\cal V}^{(2)}$. We will denote this element by $X_K$.
By definition put $X_1=(q_1,q_2)$.

Expanding the product in (\ref{g(h)-def}) we obtain an expression
\begin{equation}
\label{g_lh^l} %
g_l(h)=\frac{1}{V^{N-s}}\int dq_3\cdots dq_N \sum_{1\le
K_1<\cdots<K_l\le {\cal N}}h(X_{K_1})\cdots h(X_{K_l})
\end{equation}                                         
for every $l=1,\ldots,{\cal N}$. Introduced operators $g_l(h)$ as
well as $g(h)$ are symmetrical functions of two configurational
variables: $g_l(h)=g_l(q,q';[h])$ and $g(h)=g(q,q';[h])$ . In
contrast to $h$ and $f$ both $g(q,q';[h])$ and $g_l(q,q';[h])$
don't satisfy condition (\ref{cond_f-h}) except for
$g_1(q,q';[h])$. First term of series (\ref{g(h)}) is $g_1(h)=Bh$
and satisfies to condition (\ref{cond_f-h}). Configurational
integral (\ref{Q_N1}) takes the form
\begin{equation}
\label{Q_N2} %
Q_N=V^N(1+\overline{\overline{g}})=V^N[1+\sum_{k=2}^{\cal
N}\overline{\overline{g}}_{k}].
\end{equation}                                            
Here we used notations (\ref{h-(k)}).

Substituting definitions (\ref{g(h)-def}) and (\ref{Q_N2}) into
(\ref{h-f_1}) we write it in the form
\begin{equation}
\label{F(h-f_3)} %
(1+f)[1+\overline{\overline{g}}(h)]-[1+g(h)]=0,
\end{equation}                                            
where $f$ and $g(h)$ are functions of $X_1=(q_1,q_2)$. But value
$\overline{\overline{g}}$ doesn't depend on configurational
variables, it is a functional relative $h$. Substituting
expansions (\ref{g(h)}) and (\ref{Q_N2}) here we reduce this
equation to the form
\begin{equation}
\label{F(h-f_4)} %
f[1+\sum_{l=2}^{\cal N}
\overline{\overline{g}}_l(h)]+\sum_{l=2}^{\cal N}
[\overline{\overline{g}}_l(h)-g_l(h)]-Bh=0.
\end{equation}                                            
Here we have taken into account that $g_1(h)=Bh$ and
$\overline{\overline{g}}_1(h)=0$.

For subsequent calculation we need multilinear operators
\begin{equation}
\label{G_l} %
 G_l(y_1,\ldots,y_l)=\frac{1}{V^{N-s}}\int dq_3\cdots dq_N
\sum_{1\le K_1<\cdots<K_l\le {\cal N}}y_1(X_{K_1})\cdots
y_l(X_{K_l}).
\end{equation}                                         
These operators are linear with respect to any functional argument
$y_i$. We can consider the operator functions $g_l(h)$ as
generated by these multilinear operators $G_l$
\begin{equation}
\label{g_l-G_l} %
g_l(h)=G_l(h,\ldots,h).
\end{equation}                                         
Operators $G_l(y_1,\ldots,y_l)$ are functions of configurational
variables $\{q_1,q_2\}=X_1$. In general these functions aren't
symmetrical relative to $(q_1,q_2)$. But this isn't important
since under substituting of these operator functions into equation
(\ref{F(h-f_4)}) symmetric property will be hold automatically. In
the result we can rewrite equation (\ref{F(h-f_4)}) as
\begin{eqnarray}
 h=B^{-1}f[1+\sum_{l=2}^{\cal
N}\overline{\overline{G}}_l(h,\ldots,h)]+B^{-1}\sum_{l=2}^{\cal N}
[\overline{\overline{G}}_l(h,\ldots,h)-G_l(h,\ldots,h)].
\label{F(h-f_5)} %
\end{eqnarray}                                      

We will search a solution of this equation in the form of power
series
\begin{equation}
\label{h(f)} %
h=\sum_{k=1}^{\infty}h_k(f),
\end{equation}                                         
where $h_k(f)$ are uniform operators of order $k$ relative to $f$.
At the same time they are functions of configurational variables
$X_i$. Substituting (\ref{h(f)}) into (\ref{F(h-f_5)}) and taken
into account linearity of $G_l(y_1,\ldots,y_l)$ with respect to
any argument $y_i$ we obtain
\begin{eqnarray}
 \sum_{k=1}^{\infty}h_k(f)=B^{-1}f+B^{-1}\sum_{l=2}^{\cal
N}\sum_{j_1=1}^{\infty}\cdots\sum_{j_l=1}^{\infty}
\overline{\overline{G}}_l(h_{j_1},\ldots,h_{j_l})f\nonumber\\
\label{f/G(h)1} %
+B^{-1}\sum_{l=2}^{\cal
N}\sum_{j_1=1}^{\infty}\cdots\sum_{j_l=1}^{\infty}
[\overline{\overline{G}}_l(h_{j_1},\ldots,h_{j_l})-
G_l(h_{j_1},\ldots,h_{j_l})].
\end{eqnarray}                                      
Transform sums over $j_1,\ldots,j_l$ as follows
$$
\sum_{j_1=1}^{\infty}\cdots\sum_{j_l=1}^{\infty}=\sum_{k=l}^{\infty}\:
\sum_{j_1+\cdots+j_l=k}.
$$
Then relation (\ref{f/G(h)1}) takes the form
\begin{eqnarray}
 \sum_{k=1}^{\infty}h_k(f)=B^{-1}f+B^{-1}\sum_{l=2}^{\cal
N}\sum_{k=l}^{\infty}\sum_{j_1+\cdots+j_{\,l}=k}
\overline{\overline{G}}_l(h_{j_1},\ldots,h_{j_l})f\nonumber\\
\label{f/G(h)2} %
+B^{-1}\sum_{l=2}^{\cal
N}\sum_{k=l}^{\infty}\sum_{j_1+\cdots+j_l=k}
[\overline{\overline{G}}_l(h_{j_1},\ldots,h_{j_l})-
G_l(h_{j_1},\ldots,h_{j_l})].
\end{eqnarray}                                      

Double sum $\sum_{l=2}^{\cal N}\sum_{k=l}^{\infty}$ is transformed
as follows%
$$
\sum_{l=2}^{\cal N}\sum_{k=l}^{\infty}=\sum_{l=2}^{{\cal
N}-1}\sum_{k=l}^{{\cal N}-1}+\sum_{l=2}^{\cal N}\sum_{k={\cal
N}}^{\infty}=\sum_{k=2}^{{\cal N}-1}\sum_{l=2}^k+\sum_{k={\cal
N}}^{\infty}\sum_{l=2}^{\cal N}
$$
\mbox{or as follows}\nonumber\\
$$
\sum_{l=2}^{\cal N}\sum_{k=l}^{\infty}=\sum_{l=2}^{\cal
N}\sum_{k=l}^{\cal N}+\sum_{l=2}^{\cal N}\sum_{k={\cal
N}+1}^{\infty}=\sum_{k=2}^{\cal N}\sum_{l=2}^k+\sum_{k={\cal
N}+1}^{\infty}\sum_{l=2}^{\cal N}.
$$

Substituting these expressions into (\ref{f/G(h)2}) we get the
relation
\begin{eqnarray}
 \sum_{k=1}^{\infty}h_k(f)=B^{-1}f+B^{-1}\sum_{k=3}^{\cal
N}\sum_{l=2}^{k-1}\sum_{j_1+\cdots+j_l=k-1}
\overline{\overline{G}}_l(h_{j_1},\ldots,h_{j_l})f\nonumber\\
+B^{-1}\sum_{k={\cal N}+1}^{\infty}\sum_{l=2}^{\cal N}
\sum_{j_1+\cdots+j_l=k-1}
\overline{\overline{G}}_l(h_{j_1},\ldots,h_{j_l})f\nonumber\\
+B^{-1}\sum_{k=2}^{\cal N}\sum_{l=2}^k\sum_{j_1+\cdots+j_l=k}
[\overline{\overline{G}}_l(h_{j_1},\ldots,h_{j_l})-
G_l(h_{j_1},\ldots,h_{j_l})]\nonumber\\
+B^{-1}\sum_{k={\cal N}+1}^{\infty}\sum_{l=2}^{\cal
N}\sum_{j_1+\cdots+j_l=k}
[\overline{\overline{G}}_l(h_{j_1},\ldots,h_{j_l})-
G_l(h_{j_1},\ldots,h_{j_l})].               \label{f/G(h)3} %
\end{eqnarray}                                      
Here in the first two sums of the right-hand side we change
summation variable $k$ to $k+1$.

In this relation all sums with respect to $k$ contain expressions
of order $k$ relative to $f$. Putting terms of the same order
being equal in accordance with the theorem on uniqueness of
analytical operators \cite{Hille} we obtain the next recurrent
system for the
functions $h_k(f)$ %

\begin{eqnarray}
\label{h_1(f)}%
h_1(f)=B^{-1}f,   \\                                 
\label{h_2(f)}%
h_2(f)=B^{-1}[\overline{\overline{G}}_2(h_1,h_1)-G_2(h_1,h_1)],\\ 
h_k(f)=B^{-1}\sum_{l=2}^{k-1} \sum_{j_1+\cdots+j_l=k-1}
\overline{\overline{G}}_l(h_{j_1},\ldots,h_{j_l})f+B^{-1}\sum_{l=2}^{k} \nonumber\\
\label{h_k(f)1}%
\cdot\sum_{j_1+\cdots+j_{\,l}=k}
[\overline{\overline{G}}_l(h_{j_1},\ldots,h_{j_l})-
G_l(h_{j_1},\ldots,h_{j_l})], \qquad
3\le k\le {\cal N}, \\                                   
h_k(f)=B^{-1}\sum_{l=2}^{\cal N} \sum_{j_1+\cdots+j_l=k-1}
\overline{\overline{G}}_l(h_{j_1},\ldots,h_{j_l})f+B^{-1}
\sum_{l=2}^{\cal N} \nonumber\\
\label{h_k(f)2}%
\cdot\sum_{j_1+\cdots+j_{\,l}=k}
[\overline{\overline{G}}_l(h_{j_1},\ldots,h_{j_l})-
G_l(h_{j_1},\ldots,h_{j_l})], \qquad k\ge{\cal N}+1.
\end{eqnarray}                                      
All terms of series (\ref{h(f)}) are calculated from this system.
So the solution of the equation (\ref{F(h-f_2)}) is founded. It
satisfies additional condition (\ref{F(h0,f0)}). Convergence of
the series (\ref{h(f)}) with $h_k(f)$ being the solutions of
system (\ref{h_1(f)})-(\ref{h_k(f)2}) is proved by Cauchy-Goursat
method presented in the book \cite{Vainberg}.

As soon as $h_k$ are expressed in terms of $f$ we can get the
canonical distribution (\ref{D_N1}) in terms of $F_2$ since $f$
and $F_2$ are uniquely bounded by relation (\ref{f}).

\section{Calculation procedure for the canonical distribution
 in terms of $f$ \label{sec-phi(f)}}

Since canonical distribution (\ref{D_N1}) is a ratio of two
polynomials with respect to $h$, we see that  $D_N$ is an
analytical operator function of $h$. We have just proved that $h$
is an analytical operator function of $f$. Therefore $D_N$ is an
analytical operator function of $f$ and it can be expanded into an
absolutely convergent series relative to $f$
\begin{equation}
\label{D_N(f)} %
D_N=V^{-N}\Bigl[1+ \sum_{k=1}^{\infty}\varphi_k(f)\Bigr],
\end{equation}                                       
where $\varphi_k(f)$ is a uniform operator of order $k$
transforming function $f(q,q')$ to function
$\varphi_k(q_1,\ldots,q_N;[f])$. Taking into account definitions
of the reduced distribution functions (\ref{F_l}) and the function
$f(q,q')$ (\ref{f}) we can get relations for $\varphi_k$
\begin{eqnarray}
\label{phi-1-int} %
\frac{1}{V^{N-s}}\int\!dq_3\cdots dq_N
\varphi_1(q_1,\ldots,q_N;[f])=f(q_1,q_2),\\     
\label{phi-k-int} %
\int\!dq_3\cdots dq_N \varphi_k(q_1,\ldots,q_N;[f])=0,\qquad
k=2,3,\ldots .
\end{eqnarray}                                       
Below we construct a procedure for calculation of functions
$\varphi_1(q_1,\ldots,q_N;[f])$ in terms of $f$.

We introduce a nonlinear operator function $\lambda(h)$ by the
relation
\begin{equation}
\label{l(h)-def} %
\prod_{1\le j<k\le N}[1+h(q_j,q_k)]=1+\lambda(h).
\end{equation}                                      
This operator function is a polynomial of degree ${\cal N}$
relative to $h$. It can be written in the form
\begin{equation}
\label{l(h)} %
\lambda(h)= \sum_{l=1}^{\cal N}\lambda_l(h),
\end{equation}                                      
where $\lambda_k(h)$ are defined by relations
\begin{equation}
\label{l_s(q,..)} %
\lambda_l(q_1,\ldots,q_N;[h])=\sum_{1\le K_1<\cdots<K_l\le {\cal
N}}h(X_{K_1})\cdots h(X_{K_l}).
\end{equation}                                   
 Introduce also multilinear operators
\begin{equation}
\label{L_k} %
\Lambda_l(y_1,\ldots,y_l)\eqd \sum_{1\le K_1<\cdots<K_l\le {\cal
N}}y_1(X_{K_1})\cdots y_l(X_{K_l}).
\end{equation}                                     
It's evident that
\begin{equation}
\label{l_k=L_k} %
\lambda_l(h)=\Lambda_l(h,\ldots,h).
\end{equation}                                   

The operators introduced here are connected with the operators
$g(h)$, $g_k(h)$, and $G_k(h_1,\ldots,h_k)$ by the relations
\begin{eqnarray}
\label{g-l} %
\frac{1}{V^{N-2}}\int dq_3\cdots
dq_N\lambda(q_1,\ldots,q_N;[h])=g(q_1,q_2;[h]),\\  
\label{g-l/k} %
\frac{1}{V^{N-2}}\int dq_3\cdots
dq_N\lambda_k(q_1,\ldots,q_N;[h])=g_k(q_1,q_2;[h]),\\ 
\label{G-L/k} %
\frac{1}{V^{N-s}}\int dq_{s+1}\cdots
dq_N\Lambda_k(q_1,\ldots,q_N;[h_1,\ldots,h_k])\nonumber\\
=G_k(q_1,\ldots,q_s;[h_1,\ldots,h_k]).
\end{eqnarray}                                   
In particular for $k=1$
\begin{eqnarray}
 \frac{1}{V^{N-2}}\int dq_3\cdots
dq_N\Lambda_1(q_1,\ldots,q_N;[h])  \nonumber\\
=G_1(q_1,q_2;[h])=g_1(q_1,q_2;[h])=(Bh)(q_1,q_2).      \label{G-L/1} %
\end{eqnarray}                                   

Taking into account the expression (\ref{Q_N2}) for $Q_N$ we can
write
\begin{equation}
\label{D_N-1} %
D_N=V^{-N}\frac{1+\lambda(h)}{1+\overline{\overline{g}}(h)}.
\end{equation}                                   
Comparing it with (\ref{D_N(f)}) we get the relation
\begin{equation}
\label{phi(f)} %
\sum_{k=1}^{\infty}\varphi_k(f)=\frac{\lambda(h)-\overline{\overline{g}}(h)}
{1+\overline{\overline{g}}(h)},
\end{equation}                                   
where $h$ is the operator function of $f$ calculated in previous
section. Using here the expressions for $\lambda(h)$,
$\overline{\overline{g}}(h)$ and $h(f)$ we can transform the
right-hand side of (\ref{phi(f)}) to series with respect to $f$
and thus obtain expressions for $\varphi_k(f)$. But less awkward
transformations are obtained if we construct a recurrent system
for $\varphi_k(f)$.

Multiplying (\ref{phi(f)}) by $1+\overline{\overline{g}}(h)$ and
using (\ref{g_l-G_l}) and (\ref{l_k=L_k}) we obtain
\begin{equation}
\label{phi-G-L} %
\{1+\sum_{k=2}^{\cal
N}\overline{\overline{G}}_k(h,...,h)\}\sum_{l=1}^{\infty}\varphi_l(f)
=\sum_{k=1}^{\cal N}\Lambda_k(h,...,h)-\sum_{k=2}^{\cal
N}\overline{\overline{G}}_k(h,...,h).
\end{equation}                                   
Substitution of the expansion (\ref{h(f)}) here gives
\begin{eqnarray}
 \{1+\sum_{k=2}^{\cal N}\sum_{j_1=1}^{\infty}\cdots
\sum_{j_k=1}^{\infty} \overline{G}_k^{(0)}(h_{j_1},...,h_{j_k})\}
\sum_{l=1}^{\infty}\varphi_l(f)\nonumber\\
=\sum_{k=1}^{\cal N}\sum_{j_1=1}^{\infty}\cdots
\sum_{j_k=1}^{\infty}\Lambda_k(h_{j_1},...,h_{j_k})
-\sum_{k=2}^{\cal N}\sum_{j_1=1}^{\infty}\cdots
\sum_{j_k=1}^{\infty}\overline{G}_k^{(0)}(h_{j_1},...,h_{j_k}).
\label{phi-G-L_2} %
\end{eqnarray}                                   

Further calculation is carried out in the same way as in the
previous section. We won't make it and write a recurrent system
for $\varphi_k(f)$ straight away  %

\begin{eqnarray}
\label{phi-1} %
\varphi_1(f)=\Lambda_1(h_1),\\                 
\label{phi-2} %
\varphi_2(f)=\Lambda_1(h_2)+\Lambda_2(h_1,h_1)-
\overline{\overline{G}}_2(h_1,h_1),\\               
 \varphi_k(f)=\Lambda_1(h_k)+\sum_{l=2}^k\sum_{j_1+\cdots+j_l=k}
\{\Lambda_l(h_{j_1},...,h_{j_l})-
\overline{\overline{G}}_l(h_{j_1},...,h_{j_l})\}\nonumber\\
\label{phi-k<N+1} %
-\sum_{l=3}^k\sum_{j_1+\cdots+j_l=k}
\overline{\overline{G}}_{l-1}(h_{j_1},...,h_{j_{l-1}})
\varphi_{j_l}(f),\qquad 3\le
k\le {\cal N},\\                              
\varphi_k(f)=\Lambda_1(h_k)+\sum_{l=2}^{\cal
N}\sum_{j_1+\cdots+j_l=k} \{\Lambda_l(h_{j_1},...,h_{j_l})-
\overline{\overline{G}}_l(h_{j_1},...,h_{j_l})\}\nonumber\\
\label{phi-k>N} %
-\sum_{l=3}^{{\cal N}+1}\sum_{j_1+\cdots+j_l=k}
\overline{\overline{G}}_{l-1}(h_{j_1},...,h_{j_{l-1}})
\varphi_{j_l}(f),\qquad k\ge {\cal N}+1.
\end{eqnarray}                             

In this relations we have to use the expressions for $h_r(f)$
derived from the recurrent system (\ref{h_1(f)})--(\ref{h_k(f)2}).

For example an expression for $\varphi_1(f)$ is
\begin{equation}
\label{phi-1(f)} %
\varphi_1(q_1,\ldots,q_N;[f])=\sum_{1\le j<k\le N} f(q_j,q_k)-
(N-2)\sum_{i=1}^{ N}\overline{f}(q_i).
\end{equation}                                   
This expression coincides nominally with
$\varphi_1(q_1,\ldots,q_N;[f])$ derived in the paper
\cite{Kalinin2}. However in this formula the function $f(q,q')$
depends on both the interaction potential $u_2(q,q')$ and the
external field $u_1(q)$ whereas in \cite{Kalinin2} the latter is
absent. It is easy to show that expressions
(\ref{phi-1})--(\ref{phi-k>N}) and (\ref{phi-1(f)}) satisfy
conditions (\ref{phi-1-int}), (\ref{phi-k-int}).

\section{Set of irreducible functions containing the whole information
about system under consideration \label{set}}

The program presented here can't be realized for the reduced
one-particle distribution function because not all the theorem
conditions are held in this case. Namely the operator
$B_1=\partial F_1/\partial h(q,q')$ has a nontrivial space of
zeroes. This space consists of all functions $h(q,q')\in E_1$
satisfying a condition
\begin{equation}
\label{h_0} %
\int h(q,q')dq'=0.
\end{equation}                                            
Such operator $B$ might not have an inverse one. And the reduced
one-particle distribution function doesn't contain the whole
information about system under consideration.

At the same time there is a set of irreducible functions which is
equivalent to the reduced two-particle distribution function. This
set includes the reduced one-particle distribution function and a
pair correlation function
\bigskip %
\begin{equation}
\label{kappa} %
\varkappa(q,q')=\frac{F_2(q,q')-F_1(q)F_1(q')}{F_1(q)F_1(q')}.
\end{equation}                                        
The function $F_1(q)$ can't be expressed in terms of
$\varkappa(q,q')$  and vice versa.

Taken together they form the set which describes the system under
consideration  completely because there is a one-to-one
correspondence $F_2(q,q'){\longleftrightarrow}\{F_1(q),
\varkappa(q,q')\}$. Really on the one hand $F_2(q,q')$ is
expressed in terms of mentioned set of functions by relation
(\ref{kappa}). On the other hand $F_1(q)$ is expressed as an
integral of $F_2(q,q')$ with respect to $q'$. And
$\varkappa(q,q')\}$ is defined by relation (\ref{kappa}) in which
$F_1(q)$ is presented in terms of $F_2(q,q')$.

So the function $f(q,q')$ can be expressed in terms of the set
$\{F_1(q), \varkappa(q,q')\}$. Therefore the canonical Gibbs
distribution can be expressed in terms of this set of irreducible
functions with the help of the relations
(\ref{phi-1})--(\ref{phi-k>N}) in which functions $h_l(q,q';[f])$
must be expressed in terms of $\{F_1(q)$ and $\varkappa(q,q')\}$.
Therefore this set of irreducible functions contains the whole
information about the system under consideration.

\section{Conclusion}

Using the theorem  on implicit functions in this article it is
shown  that the reduced distribution function of order two plays a
specific role for the canonical ensemble of $N$ particles with
two-body interaction and external field. The canonical Gibbs
distribution $D_N(q_1\ldots,q_N)$ can be expressed uniquely in
terms of this function $F_2(q,q')$. From here we easily conclude
that there is a one-to-one correspondence between these two
functions. This means that the reduced distribution function $F_2$
contains information about the system under consideration as much
as the whole canonical distribution $D_N$. Reduced distribution
functions of all orders can be expressed in terms of this single
function $F_2$.

The one-particle reduced distribution functions don't satisfy  the
theorem conditions. So it is impossible to express the canonical
distribution in terms of this function $F_1$. To all appearance it
contains not all information about the system under consideration.

The set of irreducible functions $\{F_1(q),\varkappa(q,q')\}$ as
well as $F_2(q,q')$ contains the whole information about the
system under consideration.

Considered theorem provides sufficient conditions for existence
and uniqueness of inverse transformation $\{f\rightarrow h\}$.
Results obtained here are valid in some neighbourhood of
$h^{(0)}=0$, $f^{(0)}=0$. The question about size of this
neighbourhood demands special investigation.


\end{document}